\documentclass[aps,prl,twocolumn, superscriptaddress,amsmath]{revtex4}

\usepackage[latin1]{inputenc}
\usepackage{graphicx}
\usepackage{dcolumn}
\usepackage{bm}

\begin{document}

\title{Evidence for short-range antiferromagnetic fluctuations in Kondo insulating YbB$_{12}$}




\author{J.-M. Mignot}
\affiliation{Laboratoire Léon Brillouin, CEA-CNRS, CEA/Saclay, 91191 Gif sur Yvette (France)}

\author{P. A. Alekseev}
\author{K. S. Nemkovski}
\affiliation{LNSR, ISSSP, Russian Research Center ``Kurchatov Institute'', 123182 Moscow (Russia)}

\author{L.-P. Regnault}
\affiliation{DRFMC/SPSMS, CEA/Grenoble, 38054 Grenoble Cedex 9 (France)}

\author{F. Iga}
\author{T. Takabatake}
\affiliation{Department of Quantum Matter, ADSM, Hiroshima University, Higashi-Hiroshima, 739-8530 (Japan)}

\date{\today}


\begin{abstract}

The spin dynamics of mixed-valence YbB$_{12}$ has been studied by inelastic neutron scattering on a high-quality single crystal. In the Kondo-insulating regime realized at low temperature, the spectra exhibit a spin-gap structure with two sharp, dispersive, in-gap excitations at $\hbar\omega \approx 14.5$ and $\approx 20$ meV. The lower mode is shown to be associated with short-range correlations near the antiferromagnetic wave vector $\mathbf{q_{0}}=(\frac{1}{2}, \frac{1}{2}, \frac{1}{2})$. Its properties are in overall agreement with those expected for a ``spin exciton''  branch in an indirect hybridization gap semiconductor. \end{abstract}




\maketitle


Heavy-fermion compounds exhibit a whole spectrum of unconventional low-temperature behaviors,  basically reflecting the existence of a very small energy scale, of the order of a few tens of kelvin, in the electron subsystem \cite{Tsun97}. This energy scale is a hallmark of strong electron correlations and, in the metallic case, is associated with the Kondo temperature below which the heavy-quasiparticle Fermi liquid forms. In insulating compounds, on the other hand ---~so-called ``Kondo insulators'' (KI) or ``mixed-valence semiconductors'' (MVSC), such as CeNiSn, Ce$_3$Bi$_4$Pt$_3$, SmB$_6$, YbB$_{12}$, or UPtSn~--- it corresponds to the opening of a very narrow, temperature-dependent, energy gap in the electron density of states \cite{Risebg00}. The physical origin of this insulating state is still incompletely understood. It has been argued \cite {Aeppli92} that a number of aspects can be explained in terms of a one-electron band picture, with a ``hybridization gap'' forming at low temperature in the electronic density of states at the Fermi energy \cite {Zwick92}. However, there is growing evidence that strong electron-electron correlations are central to the emergence of the gap behavior, and that their effects cannot be reduced to a mere renormalization of quasiparticle states. The spin dynamics of these systems is also peculiar: in most examples studied to date, inelastic neutron scattering (INS) spectra typically exhibit a spin-gap response ($\Delta_{s} \sim 1-10$ meV) at low temperature, which seems directly related to the KI state and disappears rapidly when a single-site fluctuation regime is recovered by heating \cite{Risebg00}. Information on $\mathbf {Q}$ dependences has remained rather scarce, and to a large extent inconclusive, either because of complex anisotropy effects as in CeNiSn, or because measurements were carried out only on polycrystal samples. YbB$_{12}$ is a promising candidate for further investigations: it is an archetype KI compound \cite{Kasaya85} with a simple NaCl-type crystal structure (interpenetrating $fcc$ sublattices of Yb ions and B$_{12}$ cuboctahedra), and previous inelastic neutron scattering (INS) experiments on powder \cite{BouvetDissert, Bouvet98, Nefe99} have indicated the presence of two narrow magnetic excitations near the spin-gap edge. Early single-crystal measurements \cite{Iga99} were interpreted in terms of a single dispersive low-energy mode with high intensity along [111], but the form of the $\mathbf {Q}$ dependence was not clearly established. In this Letter, we report a detailed investigation of the low-energy spin dynamics in YbB$_{12}$ showing that there indeed exist two distinct excitations, and that the lower one has a maximum in its intensity, together with a minimum in its energy dispersion, at the wavevector  $\mathbf {q_{0}}=(\frac{1}{2}, \frac{1}{2}, \frac{1}{2})$. We ascribe this behavior to short-range antiferromagnetic (AFM) correlations and argue that it could correspond to the spin-exciton branch recently predicted to occur in the KI regime of a periodic Anderson model \cite{Risebg01}.

The sample studied consisted of two high-quality single crystals (total volume $\approx0.4$ cm$^3$) grown by the traveling-solvent floating-zone method using an image furnace with four  xenon lamps \cite{IgaXgrowth}. INS experiments were performed on the thermal-beam triple-axis spectrometer 2T at the LLB (Saclay). The sample was  oriented with a $\langle110\rangle$ crystal axis normal to the scattering plane, and cooled to temperatures between 10 and 80 K inside a closed-cycle refrigerator. Neutron spectra were recorded at fixed final energy, $E_{f}= 14.7$ meV (PG 002 monochromator, analyzer, and filter on the scattered beam), yielding an energy resolution of $\approx2$ meV  at zero energy transfer.

 \begin{figure}
	\includegraphics [width=0.75\columnwidth] {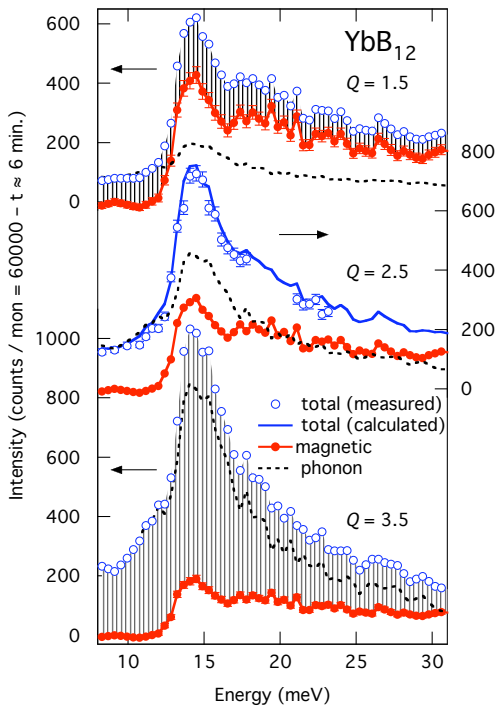}
	 \caption{\label{figPhon} (Color online) Separation of nuclear and magnetic scattering (see text) in spectra measured at $T = 11$ K for three equivalent $L$ points, $\mathbf {Q}=Q(1, 1, 1)$; hatched areas represent the phonon signal, also plotted as dashed line. }
 \end{figure}

As noted in earlier powder studies \cite{BouvetDissert,Bouvet98,Nefe99}, significant nuclear scattering from acoustic and lower optic phonon modes exists in the energy window of interest  $15 \le \hbar \omega \le 25$ meV. Special care must therefore be exerted to ensure that this signal is correctly subtracted out from the experimental spectra. To this end, we have compared data obtained at several equivalent $\mathbf {Q}$ vectors, using lattice-dynamics measurements and numerical simulations  \cite{Nemkovski04} as a guideline. The procedure is exemplified in Fig.~\ref{figPhon} for $\mathbf {Q}=\bm{\tau}+(\frac{1}{2}, \frac{1}{2}, \frac{1}{2})$, where $\bm{\tau}$ denotes a reciprocal lattice vector. Experimental data are plotted together with the calculated nuclear and magnetic components. The decomposition was made by assuming that the $Q$ dependence of the magnetic intensity follows the Yb$^{3+}$ atomic form factor, and that the phonon intensity varies proportional to $Q^2$, as expected from the model calculation for this particular set of $\mathbf {Q}$ vectors. The important check, shown in the middle frame of  Fig.~\ref{figPhon}, is that the spectrum measured at the intermediate vector  $(\frac{5}{2}, \frac{5}{2}, \frac{5}{2})$ is correctly reproduced using the partial (phonon and magnetic) contributions derived from the $\mathbf {Q}=(\frac{3}{2}, \frac{3}{2}, \frac{3}{2})$ and $(\frac{7}{2}, \frac{7}{2}, \frac{7}{2})$ spectra (Fig.~\ref{figPhon}, upper and lower frames)
\footnote{Away from the main [111] direction, the agreement is also quite good for $\mathbf {Q}=(\frac{3}{2}, \frac{3}{2}, \frac{1}{2})$ and $(\frac{1}{2}, \frac{1}{2}, \frac{5}{2})$, somewhat less so for $(\frac{1}{2}, \frac{1}{2}, \frac{3}{2})$. As a whole, the procedure appears reliable for the zone boundary $L$ point, $\mathbf {q_{0}}=(\frac{1}{2}, \frac{1}{2}, \frac{1}{2})$.}.
The contributions corresponding to the zone boundary $X$ point, $\mathbf {q}=(0, 0, 1)$, were similarly estimated from the (1, 1, 2) and (1, 1, 4) spectra, yielding a phonon signal comparable to that derived above. An average of the two was thereafter taken to represent the phonon background for all measured zone boundary spectra. On the trajectory between $(\frac{3}{2}, \frac{3}{2}, \frac{3}{2})$ and the (1, 1, 1) zone center, the phonon intensity was assumed to scale with $Q^2$
\footnote{This rather crude approximation is sufficient here because the acoustic phonon modes are weak enough in this Brillouin zone, and their dispersion is not significant in the $q$ range where $M1$ is observed. Only close to the zone center should one take into account an extra signal due to the bottom of the LO branch, which partly overlaps the upper mode $M2$.}. 
Along ($\xi, \xi, 1$), flat low-energy optic modes existing at about 24 to 28 meV throughout the Brillouin zone \cite{Nemkovski04} hamper the separation of the magnetic contribution in this energy range.

 \begin{figure}
	\includegraphics [width=0.70\columnwidth] {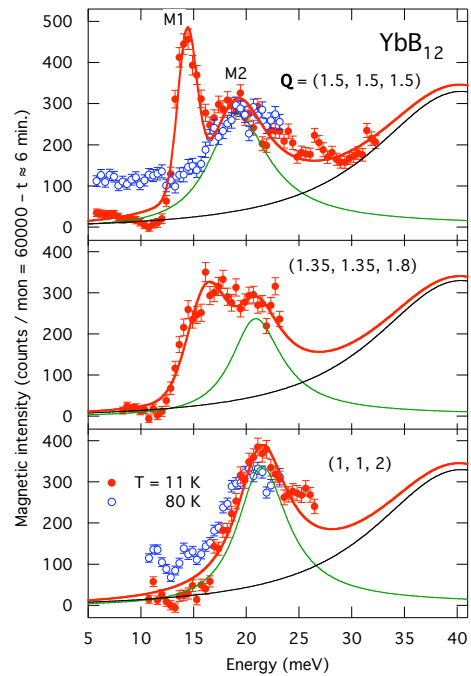}
\caption{\label{figMagnSpect}  (Color online) Magnetic excitation spectra measured for three different scattering vectors at $T=11$ and 80 K}
 \end{figure}
 
\begin{figure}
	\includegraphics [width=0.95\columnwidth] {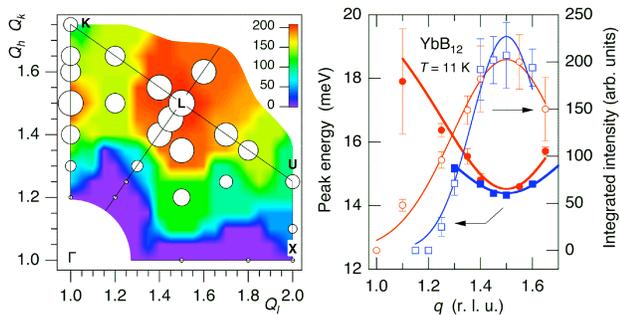}
 \caption{\label{figQdepdce}  $\mathbf {Q}$ dependence of the energies and intensities of the low-energy peak $M1$ at $T = 11$ K; left frame: intensity map over one quadrant of the Brillouin zone (circles denote $\mathbf {Q}$ vectors at which energy spectra have been measured, with diameters indicative of intensities); right frame: intensities and dispersion curves along the zone boundary (circles) and the [111] direction (squares).} 
\end{figure}

Typical magnetic excitation spectra obtained from the above treatment are displayed in Fig.~\ref{figMagnSpect}, for three different $\mathbf {Q}$ vectors on the zone boundary, $\mathbf {Q} = \mathbf {Q_{0}} \equiv (\frac{3}{2}, \frac{3}{2}, \frac{3}{2})$, (1.35, 1.35, 1.8), and (1, 1, 2), (see sketch of Brillouin zone in Fig.~\ref{figQdepdce}). In the upper frame, one notes a clear spin-gap region (no detectable signal below $\approx 10$ meV) followed by two narrow peaks, $M1$ and $M2$,  at 14.5 and 19 meV. These energies coincide with those found in the previous powder measurements, confirming that the peaks arise from the same magnetic excitations. Moving away from the [111] direction, $M1$ is strongly suppressed, while shifting to higher energies. It becomes practically undetectable in the $c^{\ast}$ direction (bottom frame). In contrast, the $M2$ peak remains visible for all three $\mathbf {Q}$ vectors, showing only a moderate increase in energy between the $L$ and $X$ points. The solid lines in Fig.~\ref{figMagnSpect} represent fits of the data using a scattering function $S_{\text{mag}}$ constituted of three spectral components, in analogy with Refs.~[\onlinecite{Nefe99, Aleks04}]


\begin{displaymath}
S_{\text{mag}}  \propto   \frac {E} {1-\exp\left(-E/k_{B}T\right)} \sum_{i=1}^{3}\chi^{\prime}_{i}(\mathbf {Q}, T) P_{i}(E, \mathbf {Q}, T).
\label{eqn1}
\end{displaymath}

The shapes of the normed spectral functions $P_{i}(E, \mathbf {Q}, T)$ were chosen Gaussian for $M1$ (nearly resolution limited) and Lorentzian for $M2$ as well as for the broad \hbox{``40-meV''} peak. The position and width of the latter component, whose maximum lies beyond the limit of our experimental energy window, were fixed (except for the form-factor $\mathbf {Q}$ dependence) at values obtained in the powder experiments \cite{Nefe99}. This is justified because the excitation results primarily from single-site processes \cite{Aleks04}.

The same procedure was applied to spectra measured at a number of scattering vectors, both inside and on the boundary of the $\bm{\tau}=(1, 1, 1)$ Brillouin zone, in order to determine the \hbox{$\mathbf {Q}$-dependence} of the lower two excitations $M1$ and $M2$. The most notable result is that the intensity of $M1$ has a pronounced maximum at the $L$ point, which also corresponds to an absolute minimum in its dispersion curve. This is demonstrated in Fig.~\ref{figQdepdce}, where the energy and integrated intensity of $M1$ are plotted as a function of the reduced $\mathbf {q}$ vector for trajectories through the $L$ point respectively parallel to [111] and along the zone boundary. The intensity drop when shifting away from $\mathbf {Q_{0}}$ is accompanied by a moderate increase in the experimental line width. An intensity map of $M1$ covering one quadrant of the projected Brillouin zone is presented in Fig.~\ref{figQdepdce}. It clearly shows the maximum centered around $\mathbf {Q_{0}}$, with a steep drop in intensity on going toward the zone center. $M1$ retains significant intensity all along $(\xi, \xi, 1)$, possibly going through a local maximum near $\mathbf {Q} = (\frac{3}{2}, \frac{3}{2}, 1)$. In contrast, it is drastically suppressed along the entire (1, 1, $\zeta$) direction.

The corresponding analysis for $M2$ is less straightforward because of the above-mentioned overlap with low-energy optic phonons. The general trend is a weaker energy dispersion in comparison with $M1$, and a much more uniform azimuthal distribution of the intensity. However, as in the case of $M1$, a pronounced reduction of the integrated signal occurs near zone center. The result is unambiguous because, in this region, $M2$ becomes narrow enough and thus can be easily separated from the phonon background. In the numerical fits, it turns out that the loss in spectral weight results mainly from a decrease in the signal line width.

\begin{figure}
	\includegraphics [width=0.95\columnwidth] {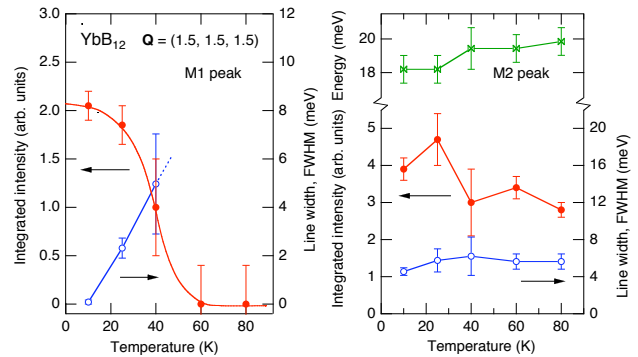}
  \caption{\label{figTmpDepdce}  Temperature dependence of fit parameters for the low-energy peaks $M1$ (left) and $M2$ (right) in the energy spectra measured at $\mathbf {Q_{0}} = (\frac{3}{2}, \frac{3}{2}, \frac{3}{2})$; lines are guides to the eye.}
 \end{figure}

With increasing temperature, one observes a complete suppression of the $M1$ excitation (see spectra for $T=80$ K in Fig.~\ref{figMagnSpect}), in agreement with the powder results \cite{BouvetDissert, Bouvet98, Nefe99}. The data also confirm the recovery of magnetic intensity, ascribed to quasi-elastic scattering, in the former spin-gap region. It should be noted that the disappearance of $M1$ takes place without appreciable shift in the excitation energy. The $T$ dependence of the parameters for $M1$ and $M2$ derived from the fit is summarized in Fig.~\ref{figTmpDepdce} for $\mathbf {Q}=\mathbf {Q_{0}}$. The $M1$ peak is steeply suppressed between 25 and 60 K, with an attending increase in its line-width (left frame). On the other hand, $M2$ (right frame) is little affected by the increase in temperature up to 80 K. 
 
The main outcome of the present measurements is the unambiguous observation of \textit{two} distinct excitations in the magnetic neutron spectrum at $T = 10$ K, exhibiting contrasted $\mathbf {Q}$ and $T$ dependences. The lower one, $M1$, is of particular interest to us because it is specific to the low-temperature KI state. The distribution of intensity over the Brillouin zone, showing a pronounced maximum near $\mathbf {Q_{0}} =\bm{\tau}_{111}+\mathbf{q_{0}}$ denotes the importance of magnetic fluctuations associated with the wave vector $\mathbf{q_{0}}=(\frac{1}{2}, \frac{1}{2}, \frac{1}{2})$. Short-range AFM correlations are  central to the physics of metallic HF systems \cite{Grewe91}, especially near quantum critical points, and there is evidence that they play a role in KI's as well \cite{Mason92, Rappoport00}. It has been reported \cite{Batko02} that some metallic heavy rare-earth \textit{R}B$_{12}$ compounds order in complex, amplitude-modulated, incommensurate structures. The paramagnetic ground state is specific to YbB$_{12}$, likely due to a Kondo spin-liquidlike phase \cite{Tsun97}. The magnetic couplings responsible for the observed AFM correlations may be affected by Fermi surface changes due to the gap formation.  The steep decrease in the intensity of $M1$ taking place between 25 and 60 K is directly correlated with the filling of the gap observed over the same temperature range in optical conductivity spectra $\sigma(\omega)$ \cite{okam98}. Basing on a parallel between the electronic structures of YbB$_{12}$ and metallic MV YbAl$_3$, Okamura et al. \cite{Okam04} have argued that the shoulder occurring below 40 meV in their low-temperature measurements of $\sigma(\omega)$ reflects the \textit{indirect} gap in the renormalized density of states expected from a periodic Anderson model. In the neutron spectra, this energy of 40 meV roughly corresponds to the position of the broad peak in the time-of-flight data \cite{Nefe99}, which disappears on heating in the same temperature region ($T \gtrsim 50$ K) as the anomaly in $\sigma(\omega)$, and we thus believe that it represents a plausible estimate of the spin gap value. If this is right, $M1$ has to be considered an \hbox{\textit{in-gap}} mode, analogous to an exciton in a conventional semiconductor.
 
This view is in line with the  model proposed by Riseborough in Ref.~\onlinecite{Risebg01}, which predicts a sharp branch of in-gap dispersive ``spin-exciton'' excitations to form, under certain conditions, in KI systems as a result of AFM exchange interactions $J(\mathbf{Q})$. This process is favored by the enhancement of the real part of the dynamical susceptibility $\chi'(\mathbf{Q},\omega)$ taking place at the spin-gap edge for $\mathbf{Q}$ vectors close to a zone boundary AFM point \footnote{In Ref.~\onlinecite{Risebg01}, the effect occurs in $(\pi, \pi, \pi)$ for a simple cubic lattice, which is relevant to SmB$_6$ but not directly to YbB$_{12}$}. The excitations are described as the continuation of an AFM paramagnon branch which softens at low temperature and undergoes strong narrowing due to the suppression of the electron-hole decay channel when it eventually falls below the threshold of the Stoner continuum. Moving toward the zone center, $J(\mathbf{Q})$ decreases and $\chi'(\mathbf{Q},\omega)$ broadens, leading to a positive dispersion, and the gradual disappearance, of the spin-exciton branch as observed in the present experiments. 

In this interpretation, the damping, and subsequent disappearance of $M1$ with increasing $T$ is a direct consequence of the gap filling evidenced by optical and transport measurements. The model implicitly assumes that the gap forms through a hybridization process taking place coherently at each $4f$ ion site, and would thus be expected to be sensitive to disorder caused by chemical substitution on the Yb sublattice. However, neutron spectra \cite{Aleks04} for the Yb$_{1-x}$Lu$_x$B$_{12}$ solid solutions indicate that the spin gap is quite robust and can be traced up to 90\% Lu dilution, which is at variance with its rapid suppression with increasing temperature (recovery of quasielastic scattering). This raises the interesting possibility that the spin gap could be driven by single-site effects (local spin-singlet state) as suggested by other authors \cite{Kasuya94, Liu01}, without necessarily requiring the existence of an insulating gap. Whether some degree of local character can be reconciled with the spin-exciton picture remains as a question for future investigations.

In summary, we have demonstrated a significant effect of AFM short-range correlations in YbB$_{12}$, leading to the formation of a sharp dispersive in-gap mode. The main properties of this excitation can be captured by a rather simple ``spin-exciton'' model, though the possibility of local effects, as inferred from substitution experiments, should also not be overlooked. Low-energy peaks have so far been reported for a few KIs such as SmB$_{6}$ \cite{Aleks94}, CeNiSn \cite{Sato95}, and now YbB$_{12}$, but failed  to be observed in a number of others \cite{Risebg00}. It is noteworthy that mechanisms reminiscent of the spin-exciton formation, but based on the existence of a superconducting rather than insulating gap, have been invoked to explain anomalous magnetic modes occurring in high-$T_c$ cuprates  (the well-known ``resonance peak'') \cite{Onufrieva02} or the HF compound UPd$_2$Al$_3$ \cite{Sato01}. The picture emerging here for an archetype KI may thus prove of interest for a broader class of ``spin-gap systems''.

\begin{acknowledgments}
We are grateful to E. V. Nefeodova, Y. Sidis, V. Lazukov, P. Riseborough, H. Okamura, T. Kasuya, and I. P. Sadikov for stimulating discussions. The work was supported by the grants N$^{\circ}$ HIII-2037.2003.2 (Grant for Leading Scientific School, Russia), 03-51-3036 (INTAS), and 13CE2002 (Grant-in-Aid for COE research of MEXT, Japan).
\end{acknowledgments}


\begin{thebibliography}{27} 	
\expandafter\ifx\csname natexlab\endcsname\relax\def\natexlab#1{#1}\fi
\expandafter\ifx\csname bibnamefont\endcsname\relax
  \def\bibnamefont#1{#1}\fi
\expandafter\ifx\csname bibfnamefont\endcsname\relax
  \def\bibfnamefont#1{#1}\fi
\expandafter\ifx\csname citenamefont\endcsname\relax
  \def\citenamefont#1{#1}\fi
\expandafter\ifx\csname url\endcsname\relax
  \def\url#1{\texttt{#1}}\fi
\expandafter\ifx\csname urlprefix\endcsname\relax\def\urlprefix{URL }\fi
\providecommand{\bibinfo}[2]{#2}
\providecommand{\eprint}[2][]{\url{#2}}

\bibitem[{\citenamefont{Tsunetsugu et~al.}(1997)\citenamefont{Tsunetsugu,
  Sigrist, and Ueda}}]{Tsun97}
\bibinfo{author}{\bibfnamefont{H.}~\bibnamefont{Tsunetsugu}},
  \bibinfo{author}{\bibfnamefont{M.}~\bibnamefont{Sigrist}}, \bibnamefont{and}
  \bibinfo{author}{\bibfnamefont{K.}~\bibnamefont{Ueda}},
  \bibinfo{journal}{Rev. Mod. Phys.} \textbf{\bibinfo{volume}{69}},
  \bibinfo{pages}{809} (\bibinfo{year}{1997}).

\bibitem[{\citenamefont{Riseborough}(2000)}]{Risebg00}
\bibinfo{author}{\bibfnamefont{P.~S.} \bibnamefont{Riseborough}},
  \bibinfo{journal}{Adv. Phys.} \textbf{\bibinfo{volume}{49}},
  \bibinfo{pages}{257} (\bibinfo{year}{2000}).

\bibitem[{\citenamefont{Aeppli and Fisk}(1992)}]{Aeppli92}
\bibinfo{author}{\bibfnamefont{G.}~\bibnamefont{Aeppli}} \bibnamefont{and}
  \bibinfo{author}{\bibfnamefont{Z.}~\bibnamefont{Fisk}},
  \bibinfo{journal}{Comments Condens. Matter Phys.}
  \textbf{\bibinfo{volume}{16}}, \bibinfo{pages}{155} (\bibinfo{year}{1992}).

\bibitem[{\citenamefont{Zwicknagl}(1992)}]{Zwick92}
\bibinfo{author}{\bibfnamefont{G.}~\bibnamefont{Zwicknagl}},
  \bibinfo{journal}{Adv. Phys.} \textbf{\bibinfo{volume}{41}},
  \bibinfo{pages}{203} (\bibinfo{year}{1992}).

\bibitem[{\citenamefont{Kasaya et~al.}(1985)\citenamefont{Kasaya, Iga,
  Takigawa, and Kasuya}}]{Kasaya85}
\bibinfo{author}{\bibfnamefont{M.}~\bibnamefont{Kasaya}}
\bibnamefont{ et al.},
  \bibinfo{journal}{J. Magn. Magn. Mater.} \textbf{\bibinfo{volume}{47\&48}},
  \bibinfo{pages}{429} (\bibinfo{year}{1985}).

\bibitem[{\citenamefont{Bouvet}(1993)}]{BouvetDissert}
\bibinfo{author}{\bibnamefont{Bouvet}}, \bibinfo{type}{PhD},
  \bibinfo{school}{University of Grenoble} (\bibinfo{year}{1993}).

\bibitem[{\citenamefont{Bouvet et~al.}(1998)\citenamefont{Bouvet, Kasuya,
  Bonnet, Regnault, Rossat-Mignod, Iga, Fåk, and Severing}}]{Bouvet98}
\bibinfo{author}{\bibfnamefont{A.}~\bibnamefont{Bouvet}}
\bibnamefont{ et al.},
  \bibinfo{journal}{J. Phys.: Condens. Matter} \textbf{\bibinfo{volume}{10}},
  \bibinfo{pages}{5667} (\bibinfo{year}{1998}).

\bibitem[{\citenamefont{Nefeodova et~al.}(1999)\citenamefont{Nefeodova,
  Alekseev, Mignot, Lazukov, Sadikov, Paderno, Shitsevalova, and
  Eccleston}}]{Nefe99}
\bibinfo{author}{\bibfnamefont{E.~V.} \bibnamefont{Nefeodova}}
\bibnamefont{ et al.},
  \bibinfo{journal}{Phys. Rev. B}
  \textbf{\bibinfo{volume}{60}}, \bibinfo{pages}{13507} (\bibinfo{year}{1999}).

\bibitem[{\citenamefont{Iga et~al.}(1999)\citenamefont{Iga, Bouvet, Regnault,
  Takabatake, Hiess, and Kasuya}}]{Iga99}
\bibinfo{author}{\bibfnamefont{F.}~\bibnamefont{Iga}}
\bibnamefont{ et al.},
  \bibinfo{journal}{J. Phys. Chem. Solids} \textbf{\bibinfo{volume}{60}},
  \bibinfo{pages}{1193} (\bibinfo{year}{1999}).

\bibitem[{\citenamefont{Riseborough}(2001)}]{Risebg01}
\bibinfo{author}{\bibfnamefont{P.~S.} \bibnamefont{Riseborough}},
  \bibinfo{journal}{J. Magn. Magn. Mater.} \textbf{\bibinfo{volume}{226-230}},
  \bibinfo{pages}{127} (\bibinfo{year}{2001}).

\bibitem[{\citenamefont{Iga et~al.}(1998)\citenamefont{Iga, Shimizu, and
  Takabatake}}]{IgaXgrowth}
\bibinfo{author}{\bibfnamefont{F.}~\bibnamefont{Iga}},
  \bibinfo{author}{\bibfnamefont{N.}~\bibnamefont{Shimizu}}, \bibnamefont{and}
  \bibinfo{author}{\bibfnamefont{T.}~\bibnamefont{Takabatake}},
  \bibinfo{journal}{J. Magn. Magn. Mater.} \textbf{\bibinfo{volume}{177-181}},
  \bibinfo{pages}{337} (\bibinfo{year}{1998}).

\bibitem[{\citenamefont{Nemkovski et~al.}()\citenamefont{Nemkovski, Alekseev,
  Mignot, and Tiden}}]{Nemkovski04}
\bibinfo{author}{\bibfnamefont{K.~S.} \bibnamefont{Nemkovski}}
\bibnamefont{ et al.},
  \bibinfo{note}{Physica Stat. Sol. (in press)}.

\bibitem[{\citenamefont{Alekseev et~al.}(2004)\citenamefont{Alekseev, Mignot,
  Nemkovski, Nefeodova, Shitsevalova, Paderno, Bewley, Eccleston, Clementyev,
  Lazukov et~al.}}]{Aleks04}
\bibinfo{author}{\bibfnamefont{P.~A.} \bibnamefont{Alekseev}}
\bibnamefont{ et al.},
  \bibinfo{journal}{J. Phys.: Condens. Matter}
  \textbf{\bibinfo{volume}{16}}, \bibinfo{pages}{2631} (\bibinfo{year}{2004}).

\bibitem[{\citenamefont{Grewe and Steglich}(1991)}]{Grewe91}
\bibinfo{author}{\bibfnamefont{N.}~\bibnamefont{Grewe}} \bibnamefont{and}
  \bibinfo{author}{\bibfnamefont{F.}~\bibnamefont{Steglich}}, in
  \emph{\bibinfo{booktitle}{Handbook of Physics and Chemistry of Rare Earths}},
  edited by \bibinfo{editor}{\bibfnamefont{J.}~\bibnamefont{Gschneider},
  \bibfnamefont{K.A.}} \bibnamefont{and}
  \bibinfo{editor}{\bibfnamefont{L.}~\bibnamefont{Eyring}}
  (\bibinfo{publisher}{Elsevier}, \bibinfo{address}{Amsterdam},
  \bibinfo{year}{1991}), vol.~\bibinfo{volume}{14}, p. \bibinfo{pages}{343}.

\bibitem[{\citenamefont{Mason et~al.}(1992)\citenamefont{Mason, Aeppli,
  Ramirez, Clausen, Broholm, Stuecheli, Bucher, and Palstra}}]{Mason92}
\bibinfo{author}{\bibfnamefont{T.~E.} \bibnamefont{Mason}}
\bibnamefont{ et al.},
  \bibinfo{journal}{Phys. Rev. Lett.} \textbf{\bibinfo{volume}{69}},
  \bibinfo{pages}{490} (\bibinfo{year}{1992}).

\bibitem[{\citenamefont{Rappoport et~al.}(2000)\citenamefont{Rappoport,
  Figueira, and Continentino}}]{Rappoport00}
\bibinfo{author}{\bibfnamefont{T.~G.} \bibnamefont{Rappoport}},
  \bibinfo{author}{\bibfnamefont{M.~S.} \bibnamefont{Figueira}}, \bibnamefont{and}
  \bibinfo{author}{\bibfnamefont{M.~A.} \bibnamefont{Continentino}}, 
  \bibinfo{journal}{Physics Letters A}
  \textbf{\bibinfo{volume}{264}}, \bibinfo{pages}{497} (\bibinfo{year}{2000}).

\bibitem[{\citenamefont{Bat'ko et~al.}(2002)\citenamefont{Bat'ko, Flachbart,
  Kohout, Mat'as, Meschke, Siemensmeyer, Schitsevalova, and Paderno}}]{Batko02}
\bibinfo{author}{\bibfnamefont{I.}~\bibnamefont{Bat'ko}}
\bibnamefont{ et al.},
  \bibinfo{journal}{Applied Physics A: Materials Science \& Processing}
  \textbf{\bibinfo{volume}{74}}, \bibinfo{pages}{S829} (\bibinfo{year}{2002}).

\bibitem[{\citenamefont{Okamura et~al.}(1998)\citenamefont{Okamura, Kimura,
  Shinozaki, Nanba, Iga, Shimizu, and Takabatake}}]{okam98}
\bibinfo{author}{\bibfnamefont{H.}~\bibnamefont{Okamura}}
\bibnamefont{ et al.},
  \bibinfo{journal}{Phys. Rev. B} \textbf{\bibinfo{volume}{58}},
  \bibinfo{pages}{R7496} (\bibinfo{year}{1998}).

\bibitem[{\citenamefont{Okamura et~al.}(2004)\citenamefont{Okamura, Michizawa,
  Matsunami, Kimura, Nanba, Ebihara, Iga, and Takabatake}}]{Okam04}
\bibinfo{author}{\bibfnamefont{H.}~\bibnamefont{Okamura}}
\bibnamefont{ et al.},
  \bibinfo{journal}{J. Magn. Magn. Mater.} \textbf{\bibinfo{volume}{272-276}},
  \bibinfo{pages}{e51} (\bibinfo{year}{2004}).

\bibitem[{\citenamefont{Okamura et~al.}(2000)\citenamefont{Okamura, Matsunami,
  Inaoka, Nanba, Kimura, Iga, Hiura, Klijn, and Takabatake}}]{Okam00}
\bibinfo{author}{\bibfnamefont{H.}~\bibnamefont{Okamura}}
\bibnamefont{ et al.},
  \bibinfo{journal}{Phys. Rev. B} \textbf{\bibinfo{volume}{62}},
  \bibinfo{pages}{R13265} (\bibinfo{year}{2000}).

\bibitem[{\citenamefont{Kasuya}(1994)}]{Kasuya94}
\bibinfo{author}{\bibfnamefont{T.}~\bibnamefont{Kasuya}},
  \bibinfo{journal}{Europhys. Lett.} \textbf{\bibinfo{volume}{26}},
  \bibinfo{pages}{277} (\bibinfo{year}{1994}).

\bibitem[{\citenamefont{Liu}(2001)}]{Liu01}
\bibinfo{author}{\bibfnamefont{S.~H.} \bibnamefont{Liu}},
  \bibinfo{journal}{Phys. Rev. B} \textbf{\bibinfo{volume}{63}},
  \bibinfo{pages}{115108} (\bibinfo{year}{2001}).

\bibitem[{\citenamefont{Alekseev et~al.}(1995)\citenamefont{Alekseev, Mignot,
  Rossat-Mignod, Lazukov, Sadikov, Konovalova, and Paderno}}]{Aleks94}
\bibinfo{author}{\bibfnamefont{P.}~\bibnamefont{Alekseev}}
\bibnamefont{ et al.},
  \bibinfo{journal}{J. Phys.: Condens. Matter} \textbf{\bibinfo{volume}{7}},
  \bibinfo{pages}{289} (\bibinfo{year}{1995}).

\bibitem[{\citenamefont{Sato et~al.}(1995)\citenamefont{Sato, Kadowaki,
  Yoshizawa, Ekino, Takabatake, Fujii, Regnault, and Isikawa}}]{Sato95}
\bibinfo{author}{\bibfnamefont{T.~J.} \bibnamefont{Sato}}
\bibnamefont{ et al.},
  \bibinfo{journal}{J. Phys.: Condens. Matter} \textbf{\bibinfo{volume}{7}},
  \bibinfo{pages}{8009} (\bibinfo{year}{1995}).

\bibitem[{\citenamefont{Onufrieva and Pfeuty}(2002)}]{Onufrieva02}
\bibinfo{author}{\bibfnamefont{F.}~\bibnamefont{Onufrieva}} \bibnamefont{and}
  \bibinfo{author}{\bibfnamefont{P.}~\bibnamefont{Pfeuty}},
  \bibinfo{journal}{Phys. Rev. B} \textbf{\bibinfo{volume}{65}},
  \bibinfo{pages}{054515/1} (\bibinfo{year}{2002}).

\bibitem[{\citenamefont{Sato et~al.}(2001)\citenamefont{Sato, Aso, Miyake,
  Shiina, Thalmeier, Vareloglannis, Gelbel, Steglich, Fulde, and
  Komatsubara}}]{Sato01}
\bibinfo{author}{\bibfnamefont{N.~K.} \bibnamefont{Sato}}
\bibnamefont{ et al.},
  \bibinfo{journal}{Nature} \textbf{\bibinfo{volume}{410}},
  \bibinfo{pages}{340} (\bibinfo{year}{2001}).


\end{thebibliography}

\end{document}